\documentclass[english,aps,manuscript]{revtex4}
\usepackage[]{fontenc}
\usepackage[latin9]{inputenc}
\usepackage{amsmath}
\usepackage{graphicx}
\usepackage{amssymb}
\usepackage{esint}

\makeatletter

\providecommand{\LyX}{L\kern-.1667em\lower.25em\hbox{Y}\kern-.125emX\@}

\makeatother

\usepackage{babel}

\begin{document}

\preprint{}

\title{Black hole complementarity from AdS/CFT}

\author{David A. Lowe}

\email{lowe@brown.edu}

\affiliation{Physics Department, Brown University, Providence, RI, 02912, USA}
\begin{abstract}
We study a simple version of the AdS/CFT (anti-de Sitter spacetime/Conformal
Field Theory) correspondence, where operators have integer conformal
dimensions. In this model, bulk causality follows from boundary analyticity,
even in nontrivial black hole backgrounds that break the underlying
conformal symmetry. This allows a natural set of quasi-local bulk
observables to be constructed. Estimates of finite central charge
corrections to semiclassical correlators are made. These corrections
are used to determine the regime of validity of effective field theory
in the bulk spacetime. The results are consistent with black hole
complementarity.
\end{abstract}
\maketitle

\section{introduction}

The Anti-de Sitter/Conformal Field Theory correspondence \citep{Maldacena:1997re}
provides us with one of the most promising approaches to developing
a nonperturbative formulation of string theory. In the present work,
our goal is to use known properties of conformal field theory to learn
about quantum gravity in an asymptotically anti-de Sitter background.
One of the necessary steps toward this goal is to learn as much as
possible about the mapping between conformal field theory operators
and bulk observables. In particular, we would like to understand how
bulk causality follows from properties of the boundary theory in a
semiclassical limit, and how quantum effects correct these results. 

The black hole information paradox \citep{Hawking:1976ra} arises
as an incompatibility between unitary quantum theory, and the purely
thermal emission of Hawking radiation that appears in a semiclassical
approximation. With the advent of AdS/CFT, we believe that the underlying
quantum theory is a unitary conformal field theory \citep{Lowe:1999pk,Maldacena:2001kr}.
The essence of the paradox then becomes understanding why effective
field theory breaks down in a region of spacetime where curvatures
are small \citep{Lowe:1995ac} (for example a region that includes
points inside the horizon, and points outside in a region with a significant
amount of Hawking radiation). Progress on the questions raised above
can thus lead to a resolution of the paradox.

In a series of papers, the mapping from boundary CFT correlators to
on-shell bulk correlators has been constructed via a kind of inverse
LSZ \citep{H.1955}method \citep{Hamilton:2005ju,Hamilton:2006az,Hamilton:2006fh,Hamilton:2007wj,Tolfree:2008am}
for Lorentzian signature anti-de Sitter spacetime, and the results
have been generalized to the three-dimensional black hole of BTZ \citep{Banados:1992wn}
in \citep{Hamilton:2006fh,Hamilton:2007wj}. It is clear from this
construction that the large central charge limit of the boundary correlators
reproduces the expected semiclassical bulk correlators. This was shown
in the free limit in \citep{Hamilton:2006fh}. CFT correlators have
been extracted from the interacting supergravity theory in \citep{D'Hoker:1999ea,D'Hoker:2002aw}
(i.e. perturbatively in $1/N$ in the case of 4d $SU(N)$ CFT).

In the present work, the first goal is to build on the results of
\citep{Hamilton:2005ju,Hamilton:2006az,Hamilton:2006fh,Hamilton:2007wj}
and develop a model where bulk causality can be derived for a more
general class of states in the CFT, or asymptotically AdS backgrounds
on the gravity side. To make progress on this, we make the simplifying
assumption that the conformal weights in the CFT take integer values.
Holomorphic CFT's and the extremal CFT's described in \citep{Witten:2007kt}
provide examples of this type. These CFT's have retarded boundary
propagators that are non-vanishing only at light-like separations.
Following the prescription of \citep{Hamilton:2005ju,Hamilton:2006az,Hamilton:2006fh,Hamilton:2007wj}
we extract a straightforward geometric picture that determines when
a commutator of quasi-local bulk observables is non-vanishing. At
large central charge this matches with bulk causality, even for the
BTZ black hole background.

The next goal is to take some steps toward understanding nonperturbative
corrections from the bulk perspective. This immediately raises a number
of questions of principle. A finite central charge CFT is not expected
to have an exact continuum bulk spacetime interpretation. However
for large central charge, $c$, we expect to have a host of corrections
perturbative in $1/c$ that will modify the boundary-bulk map of \citep{Hamilton:2005ju,Hamilton:2006az,Hamilton:2006fh,Hamilton:2007wj}.
In general we do not expect these type of corrections to drastically
modify the picture bulk causality that emerges from the above considerations. 

A qualitative change comes when we consider effects nonperturbative
in $1/c$. A useful discussion of these effects on the boundary CFT
correlators arising from a black hole background can be found in \citep{Birmingham:2002ph,Barbon:2003aq,Barbon:2004ce,Barbon:2005jr,Kleban:2004rx,Solodukhin:2005qy}.
The upshot of this discussion is that a thermal correlator decays
exponentially, with a timescale of order the inverse temperature,
until it reaches a relative magnitude of order $e^{-S_{bh}}$, where
$S_{bh}$ is the Bekenstein-Hawking entropy of the black hole. Beyond
this time, nonperturbative effects dominate the boundary two-point
function. This can be incorporated into the boundary-bulk mapping,
as described in preliminary form in \citep{Hamilton:2007wj}. The
idea is simply to place a time cutoff on the boundary time integrals
needed to construct quasi-local bulk operators. The regime of validity
of an effective field theory based on these operators may then be
extracted simply by asking when the results start to significantly
differ from the semiclassical results. 

One can imagine extending this construction to states more general
than black holes. An issue that arises is defining when a family of
states in the CFT corresponds to a smooth region of bulk spacetime.
Certainly we expect one needs to include corrections to the {}``smearing
functions'' that appear in the mapping to take into account back-reaction
of the state on the bulk spacetime. However finite $c$ corrections
can still lead to no semiclassical bulk description. It seems the
natural way to proceed is to try to optimize corrections to the construction,
and try to find finite regions of bulk spacetime for which a large
set of low energy observables can be obtained, and their correlators
reproduced to reasonable accuracy by a local bulk effective action.
While we can expect the local effective action to respect general
covariance, we do not expect this of the finite $c$ corrections.
In this sense we define continuous bulk spacetime by a large set of
correlators that can be extracted from the CFT, and reproduced to
some reasonable accuracy by a local effective action. The local effective
action then gives a complete summary of the, necessarily approximate,
bulk interpretation of the physics. In general, one may need a number
of distinct effective actions defined on overlapping regions of bulk
spacetime to maximally extend the bulk description. However there
is no guarantee that this set of effective actions can be replaced
by a single local action, as would be expected by general covariance.

\section{Review of Bulk-Boundary Mapping for Pure AdS}

\subsection{SUGRA two-point function of massive scalar field}

For simplicity, we will restrict our considerations to bulk observables
corresponding to massive scalar fields. In general we have for the
bulk Wightman function in $AdS_{d}$ \citep{Burgess:1984ti,D'Hoker:2002aw,Dohse:2007by}
\[
G(x,x')=\frac{R^{2-d}\Gamma(\Delta)}{2^{\Delta+1}\pi^{(d-1)/2}\Gamma(\Delta-\frac{d-3}{2})}\sigma^{-\Delta}\,_{2}F_{1}(\frac{\Delta}{2},\frac{\Delta+1}{2},\Delta;\frac{1}{\sigma^{2}})\,,\]
where \[
\Delta=\frac{d-1}{2}+\sqrt{\left(\frac{d-1}{2}\right)^{2}+m^{2}R^{2}}\,,\]
and \[
\sigma=\frac{X_{\mu}Y_{\nu}\eta^{\mu\nu}}{R^{2}}\,,\]
where $\eta^{\mu\nu}$ is the $d+1$-dimensional Minkowski metric.
The AdS spacetime is realized as an embedding where \[
X_{\mu}X_{\nu}\eta^{\mu\nu}=(\vec{X})^{2}-(X^{0})^{2}-(X^{d})^{2}=-R^{2}\,.\]
In global coordinates\[
ds^{2}=\frac{R^{2}}{\cos^{2}\rho}\left(-d\tau^{2}+d\rho^{2}+\sin^{2}\rho d\Omega_{d-2}^{2}\right)\,,\]
we have the isometry invariant distance functions\[
\sigma(x,x')=\frac{\cos(\tau-\tau')-\sin\rho\sin\rho'\cos(\Omega-\Omega')}{\cos\rho\cos\rho'}\,,\]
with $\Omega-\Omega'$ the angular separation on the sphere. The operator
ordering is taken care of with a $\tau\to\tau-i\epsilon$ prescription
. This expression is correct for global AdS. To generalize it to the
covering space of AdS, we need to take into account that $\sigma$
changes sign as one moves in the time direction. To do this we introduce
the idea of a winding number, defined as
\begin{itemize}
\item $n(x,y)=0$ if $x$ can be continuously deformed to $y$ without changing
the sign of $\sigma$.
\item $\Delta n(x,y)=1$ every time $\sigma$ changes sign, for $x$ to
the future of $y$. 
\end{itemize}
Then the full expression, valid on the covering space of AdS is \[
G=e^{i\pi n\Delta}G_{n=0}\,,\]
which is obtained by analytically continuing \ref{eq:ganal}.

This can be simplified for $AdS_{3}$ to this algebraic function of
the invariant distance \begin{equation}
G(x,x')=\frac{\sigma^{\Delta-2}}{4\pi R}\frac{\left(1-\sqrt{1-\sigma^{-2}}\right)^{\Delta-1}}{\sqrt{1-\sigma^{-2}}}\,.\label{eq:ganal}\end{equation}

\subsection{Reproducing using smearing functions}

In this section we will show how \eqref{eq:ganal} can be reproduced,
following the prescription of \citep{Hamilton:2005ju,Hamilton:2006az,Hamilton:2006fh,Hamilton:2007wj}.
We will use this as an example to more fully explain the impact of
the $i\epsilon$ prescription on this construction. It is more convenient
to work in Poincare coordinates\[
ds^{2}=\frac{R^{2}}{Z^{2}}\left(-dT^{2}+dX^{2}+dZ^{2}\right)\,,\]
and we assume both bulk points are in the same coordinate patch. In
these coordinates, the invariant distance function takes the form\[
\sigma(x,x')=\frac{-T^{2}+X^{2}+Z_{1}^{2}+Z_{2}^{2}}{2Z_{1}Z_{2}}\,.\]
We begin with eqn. (16) of \citep{Hamilton:2006fh}. This expresses
an on-shell local bulk operator as a boundary operator, smeared over
an analytic continuation of the boundary manifold \begin{equation}
\phi(T,X,Z)=\frac{\Delta-1}{\pi}\int_{T'^{2}+Y'^{2}<Z^{2}}dT'dY'\left(\frac{Z^{2}-T'^{2}-Y'^{2}}{Z}\right)^{\Delta-2}\phi_{0}(T+T',X+iY')\,.\label{eq:hamil}\end{equation}
With the boundary correlator (correcting a typo in \citep{Hamilton:2006fh})
\begin{equation}
\left\langle \phi_{0}(T,X)\phi_{0}(0,0)\right\rangle _{CFT}=\frac{1}{2\pi R}\frac{1}{(X^{2}-T^{2})^{\Delta}}\,,\label{eq:bcol}\end{equation}
we obtain

\begin{eqnarray}
\left\langle \phi(T,X,Z_{1})\phi(0,0,Z_{2})\right\rangle  & = & -\frac{(\Delta-1)^{2}}{2\pi^{3}R}\int_{0}^{Z_{1}}dr_{1}\int_{0}^{Z_{2}}dr_{2}\oint_{C_{1}}dz_{1}\oint_{C_{2}}dz_{2}\label{eq:integral}\\
 &  & \frac{r_{1}r_{2}}{z_{1}z_{2}}\left(\frac{Z_{1}^{2}-r_{1}^{2}}{Z_{1}}\right)^{\Delta-2}\left(\frac{Z_{2}^{2}-r_{2}^{2}}{Z_{2}}\right)^{\Delta-2}\nonumber \\
 &  & \times\frac{1}{\left(\left(X-T+i\epsilon-r_{1}/z_{1}+r_{2}/z_{2}\right)\left(X+T-i\epsilon+r_{1}z_{1}-r_{2}z_{2}\right)\right)^{\Delta}}\,,\nonumber \end{eqnarray}
having performed a change of variables to $T'+iY'=rz$. 

We take $\Delta$ to be integer valued, to avoid issues with branch
cuts of the integrand. For scalar fields, integer conformal dimensions
are guaranteed with sufficient supersymmetry. Without supersymmetry
purely integer conformal dimensions have arisen in the proposal for
pure gravity based on an extremal conformal field theory \citep{Witten:2007kt}.
At the end of the day, one might try to analytically continue in $\Delta$,
as has been discussed in \citep{Lee:1998bxa,D'Hoker:1999ea,D'Hoker:2002aw}.
However we will not consider this approach in the present work.

The contours of integration $C_{1},C_{2}$ must be handled with some
care, and are defined as follows. The equation (\ref{eq:hamil}) can
be used as is, provided the two integration patches on the boundary
are non-overlapping. This means the contours $C_{1}$ and $C_{2}$
follow the unit circle. The $i\epsilon$ prescription then ensures
the resulting integrals are well-defined for all choices of bulk point
via analytic continuation in $ $$ $$T,X,Z_{1}$ or $Z_{2}$. This
implies the contours must be deformed as the singularities of the
integrand move, to avoid crossing.%
\footnote{With one point on the boundary, a simple prescription was given in
\citep{Hamilton:2006fh} involving excising poles in the upper-half
$z$-plane. For general points, this is no longer applicable. %
}

With this prescription, the integral \eqref{eq:integral} reproduces
the expression \eqref{eq:ganal}, as we see from the following argument.
First we note that it is straightforward to check for particular values
of $\Delta$ and general bulk points, which we have checked for $\Delta$
ranging from 2 through 10. For general integer values of $\Delta$
we have also checked agreement when one bulk point approaches the
boundary (generalizing a calculation of \citep{Hamilton:2006fh})
and when two bulk points coincide. These calculations are shown in
appendix A and B. With the $i\epsilon$ prescription, the integral
is analytic in the coordinates, so because it agrees at the singular
points, and at the boundary, it will agree for general bulk points.

\section{Bulk causality from boundary analyticity}

\subsection{Pure AdS Spacetime\label{sub:Pure-AdS-Spacetime}}

The vacuum expectation value of the commutator of boundary fields
takes a particularly simple form when $\Delta$ is integer-valued\[
\left\langle \left[\phi_{0}(x),\phi_{0}(x')\right]\right\rangle =0\,,\]
unless $x$ and $x'$ are light-like separated on the boundary. This
follows straightforwardly from the form \eqref{eq:bcol}, which is
exact due to conformal invariance. This leads to a simple geometric
picture of the commutator of two on-shell bulk operators when we use
the boundary representation \eqref{eq:hamil}. 

Consider the expectation value of the commutator of two fields in
pure $AdS_{3}$ expressed as a difference between two integrals of
the form \eqref{eq:integral}, with $i\epsilon$'s of differing signs.
Nonvanishing contributions will arise only from the singularities
of the integrand, which typically lead to branch cuts in the integral.
This may be analyzed using the standard method of Landau equations,
as explained for example \citep{R.J.1966}. This boils down to finding
the regions in parameter space where the singularities of the integrand
pinch the contour of integration, as $\epsilon\to0$. 

To proceed, one defines a comparison function with singularities at
the same position as the integral under study \eqref{eq:integral}\[
F(X,T,Z_{1},Z_{2})=\int_{H}\prod_{i=1}^{4}dw_{i}\frac{1}{S_{1}S_{2}}\,\]
with analytic functions\[
S_{1}=X+T+r_{1}z_{1}-r_{2}z_{2},\qquad S_{2}=X-T-r_{1}/z_{1}+r_{2}/z_{2}\]
and\[
w_{1}=z_{1},\quad w_{2}=z_{2},\quad w_{3}=r_{1},\quad w_{4}=r_{2}\]
and the integration hypersurface $H$ defined by the product of the
two disc regions. For the purposes of this analysis the fixed singularities
at $r_{1}=0$, $r_{2}=0$, $r_{1}=Z_{1}$, $r_{2}=Z_{2}$ can be neglected,
provided $\Delta\geq2$. Using a Feynman parameter we can then write\[
F(X,T,Z_{1},Z_{2})=\int_{H}dz_{1}dz_{2}dr_{1}dr_{2}\int d\alpha_{1}d\alpha_{2}\frac{\delta(\alpha_{1}+\alpha_{2}-1)}{(\alpha_{1}S_{1}+\alpha_{2}S_{2})^{2}}\,.\]
The conditions for a generic singularity can then be expressed as\[
S_{1}=S_{2}=0,\quad\mathrm{and}\quad\alpha_{1}\frac{\partial S_{1}}{\partial w_{i}}+\alpha_{2}\frac{\partial S_{2}}{\partial w_{i}}=0\quad\forall i\quad\mathrm{and\ for\ some}\quad\alpha_{1},\alpha_{2}\,.\]
There exist no solutions for both $\alpha_{1}\neq0$ and $\alpha_{2}\neq0$
so pinching cannot happen at some generic point on the interior of
$H$. 

Another possibility is that the surface defined by $S_{1}=0$ develops
a conical singularity, which then traps the integration surface $H$.
This happens when \[
S_{1}=0,\quad\frac{\partial S_{1}}{\partial w_{i}}=0,\quad\forall i\,.\]
However this only has solutions at the endpoints when $r_{1}=0$ and
$r_{2}=0$. We will come to this case next. A similar analysis for
$S_{2}$ shows conical trapping does not occur in this case.

To treat the case when singularities appear on the edge of the integration
region $H$, we introduce additional analytic functions $\tilde{S_{r}}$
so that $\tilde{S}_{r}=0$ defines the boundary. In the case at hand,
we have a rectangle in the $(r_{1},r_{2})$-plane. Let us begin by
considering the case that a singularity appears on an edge, away from
a corner. Take for example $\tilde{S}_{1}=r_{1}-Z_{1}$. The comparison
function may be generalized to include an additional $\tilde{S}_{1}$
factor, along with an additional Feynman parameter $\tilde{\alpha}_{1}$.
The different cases considered above may then be reduced to solving
the equations \citep{R.J.1966}\begin{equation}
\alpha_{k}S_{k}=0,\quad\forall k\quad\mathrm{and\ }\tilde{\alpha}_{1}\tilde{S}_{1}=0,\ and\ \frac{\partial}{\partial w_{i}}\left(\tilde{\alpha}_{1}\tilde{S}_{1}+\sum_{k}\alpha_{k}S_{k}\right)=0\quad\forall i\,.\label{eq:landau}\end{equation}
It is then easy to see that no solution is possible, except at the
corners of the rectangle.

\begin{figure}
\includegraphics{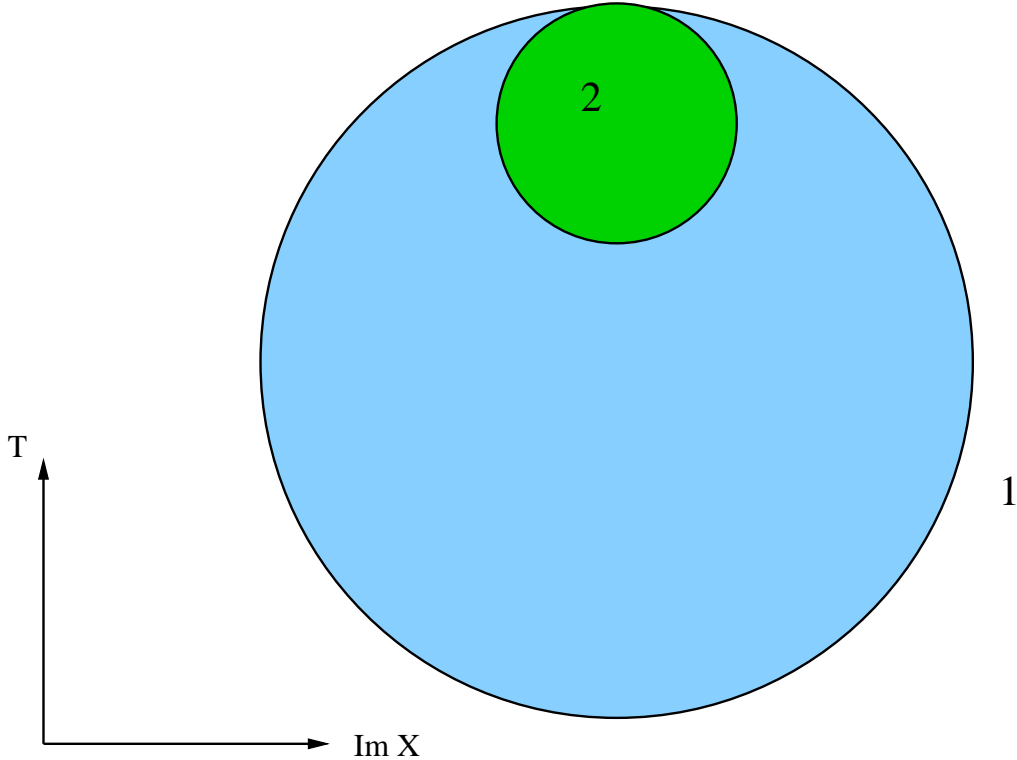}

\caption{Regions of integration on the analytic continuation of the boundary.
The commutator becomes non-trivial as soon as one disc edge touches
(is light-like separated from) the other edge. This corresponds to
bulk time-like separation, with $-1<\sigma<1.$\label{fig:Regions-of-integration}}

\end{figure}

\begin{figure}
\includegraphics{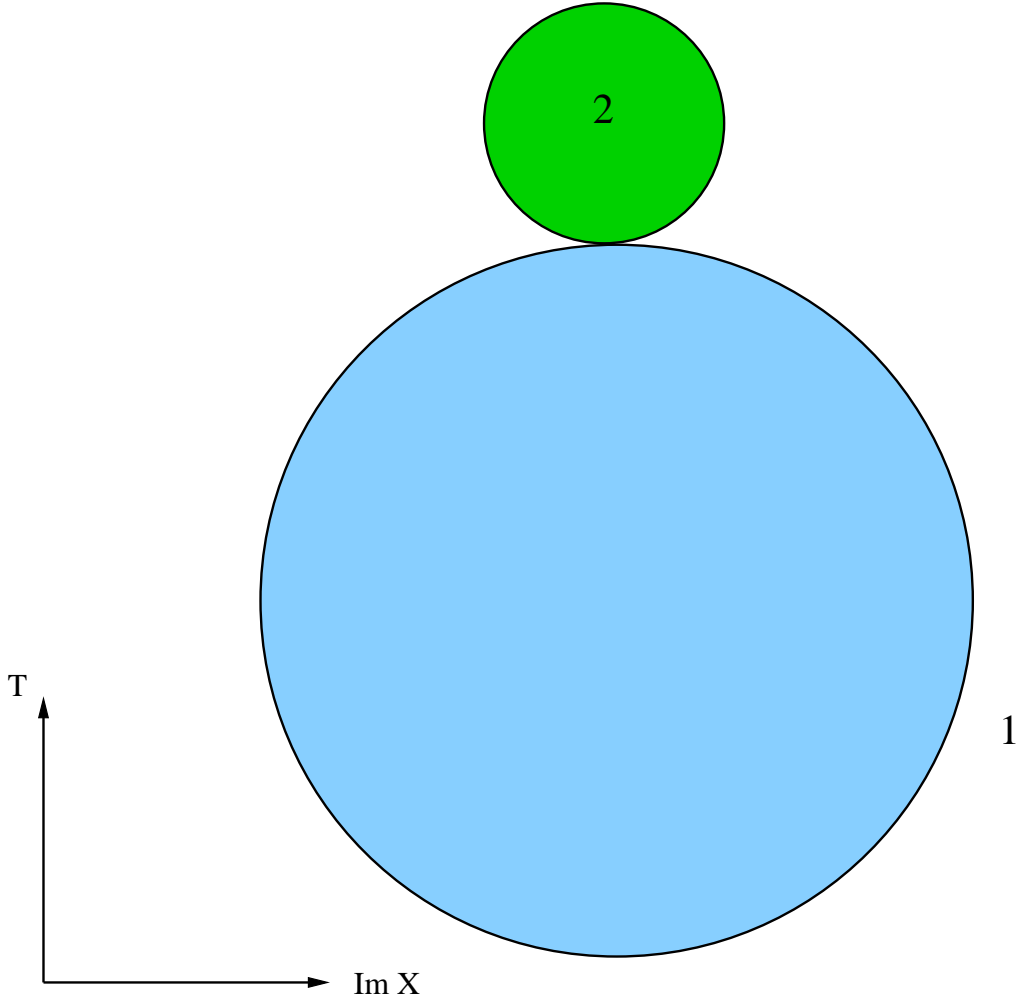}

\caption{The commutator vanishes as one disc moves outside the light-cone of
the other. This corresponds to bulk points that cannot be connected
by geodesics, $\sigma<-1$.\label{fig:The-commutator-vanishes}}

\end{figure}

To study the corners, we will use two $\tilde{S}_{r}$ factors in
our comparison function. At first sight, it appears that singularities
might appear from the $r_{1}\to0$ or $r_{2}\to0$ region. However
this is an artifact of our coordinate choice (see below \eqref{eq:integral}),
as can be easily checked by switching to Cartesian coordinates. Therefore
we need only examine the corner $r_{1}=Z_{1}$ and $r_{2}=Z_{2}$.
Solving the equations \eqref{eq:landau} yields two nontrivial solutions\begin{eqnarray*}
T^{2}-X^{2} & = & (Z_{1}-Z_{2})^{2},\quad z_{1}=z_{2}\\
T^{2}-X^{2} & = & (Z_{1}+Z_{2})^{2},\quad z_{1}=-z_{2}\end{eqnarray*}
The first case occurs when a point on the edge of disc 1 becomes light-like
separated from the edge of disc 2 as shown in figure \ref{fig:Regions-of-integration}.
The second case occurs after disc 1 has moved outside the light-cone
of disc 2, and only one point remains light-like separated, as shown
in figure \ref{fig:The-commutator-vanishes}. 

Having identified the possible positions of the singularities, we
need to check whether these actually correspond to pinches of the
hypersurface $H$ under the $T\to T-i\epsilon$ deformation, or whether
the singularities harmlessly coalesce as $\epsilon\to0$. For both
cases we find indeed the contour is pinched as $\epsilon\to0$ and
is responsible for a branch cut in the integral \eqref{eq:integral}.

This then gives a simple geometric picture of how the boundary theory
encodes bulk causality. Boundary causality guarantees a vanishing
commutator at spacelike boundary separations. Moreover for integer
$\Delta$ the commutator will also vanish at timelike separations,
and will only be non-vanishing for light-like separations. The bulk
radial coordinate is encoded in the size of the boundary disc. Because
the only non-vanishing contribution comes from a disc edge we reproduce
the expected non-vanishing of the commutator at bulk timelike separations.

\begin{figure}
\includegraphics{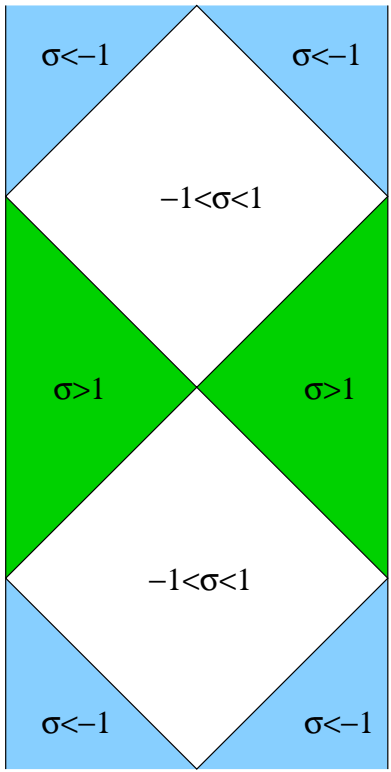}

\caption{The white regions of the AdS Penrose diagram indicate where the commutator
is non-vanishing.\label{fig:The-white-regions}}

\end{figure}

One might be puzzled by the vanishing of the commutator at bulk timelike
separations, when the discs no longer intersect. This corresponds
to the case $\sigma<-1$. As is clear from the Wightman function \eqref{eq:ganal}
the commutator will vanish in this case. Geometrically, this corresponds
to bulk points that cannot be connected by timelike geodesics. These
regions appear because the negative cosmological constant {}``repels''
timelike geodesics from spacelike infinity. The commutator is only
non-vanishing at timelike separations within a sequence of causal
diamonds, corresponding to $-1\leq\sigma\leq1$, as shown in figure
\ref{fig:The-white-regions}.

\subsection{Finite temperature and the BTZ Black Hole}

Now we wish to consider the case of the BTZ black hole. It is most
convenient to now switch to Rindler coordinates to describe $AdS_{3}$,
\begin{equation}
ds^{2}=\frac{R^{2}}{r^{2}-r_{+}^{2}}dr^{2}-\frac{r^{2}-r_{+}^{2}}{R^{2}}dt^{2}+r^{2}d\phi^{2}\,,\label{eq:btzmetric}\end{equation}
where $\phi\in\mathbb{R}$ for pure AdS. We obtain the BTZ black hole
(with vanishing angular momentum) simply by periodically identifying
$\phi\sim\phi+2\pi$. The parameter $R$ is the radius of curvature
of the AdS space, while $r_{+}$ represents the position of the Rindler
(or BTZ) horizon. Correlation functions in the Hartle-Hawking vacuum
are obtained by considering the Euclidean geometry periodically identifying
in imaginary time $t\sim t+i\beta$, with $\beta=1/T_{H}=2\pi R^{2}/r_{+}$
the inverse Hawking temperature. %
\begin{figure}
\includegraphics{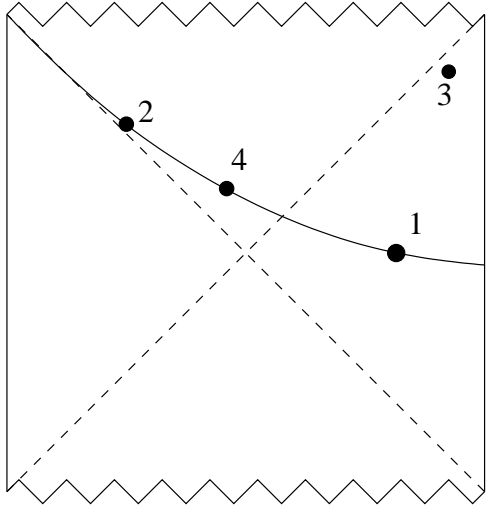}

\caption{The Penrose diagram for the BTZ black hole\label{fig:The-Penrose-diagram}.}

\end{figure}

Let us briefly recall the results of \citep{Hamilton:2006fh} on the
bulk-boundary mapping in this case. For spacetime points in the right
Rindler wedge (i.e. the right triangle in figure \ref{fig:The-Penrose-diagram}),
the map from boundary to bulk operators can be written in the form\begin{equation}
\phi(t,r,\phi)=\frac{(\Delta-1)2^{\Delta-2}}{\pi R^{3}}\int_{spacelike}dx\, dy\,\lim_{r'\to\infty}\left(\sigma/r'\right)^{\Delta-2}\phi_{0}^{R}(t+x,\phi+iy)\label{eq:bulkbound}\end{equation}
where $\phi$ is the bulk operator, $\phi_{0}^{R}$ is the boundary
operator associated with the right Rindler patch, and $\sigma$ is
the analytic continuation of the invariant distance \[
\sigma(t,r,\phi|t+x,r',\phi+iy)=\frac{rr'}{r_{+}^{2}}\left[\cos\frac{r_{+}y}{R}\mp\left(\frac{r_{+}^{2}}{r^{2}}-1\right)^{1/2}\sinh\frac{r_{+}x}{R^{2}}\right],\]
with signs determined by the right/left Rindler patch. The relation
\eqref{eq:bulkbound} may also be generalized to points inside the
horizon, where now a piece coming from the boundary of the left Rindler
patch is also needed\begin{eqnarray}
\phi(t,r,\phi) & = & \frac{(\Delta-1)2^{\Delta-2}}{\pi R^{3}}\left[\int_{\sigma>0}dx\, dy\,\lim_{r'\to\infty}\left(\sigma/r'\right)^{\Delta-2}\phi_{0}^{R}(t+x,\phi+iy)\right.\nonumber \\
 &  & \left.+\int_{\sigma<0}dx\, dy\,\lim_{r'\to\infty}\left(-\sigma/r'\right)^{\Delta-2}(-1)^{\Delta}\phi_{0}^{L}(t+x,\phi+iy)\right]\label{eq:bulkrtlf}\end{eqnarray}
One can choose to represent the contribution from the left Rindler
patch as a contour integral in the complex $t$-plane of the CFT corresponding
to the right boundary, as discussed in \citep{Kraus:2002iv,Hamilton:2006fh}.
This is achieved using the relation\begin{equation}
\phi_{0}^{L}(t,\phi)=\phi_{0}^{R}(t+i\pi R^{2}/r_{+},\phi)\label{eq:antipode}\end{equation}
 relating boundary fields in the left and right Rindler patches.

The identifications used in constructing the BTZ geometry break the
Lorentz invariance of the boundary conformal field theory. Nevertheless,
since the correlators may be defined as weighted averages in the original
conformal field theory\begin{equation}
\left\langle \phi(x)\phi(x')\right\rangle _{T}=\sum_{M}\left\langle M\vert\phi(x)\phi(x')\vert M\right\rangle \frac{e^{-\beta E_{M}}}{Z(T)}\label{eq:therm}\end{equation}
with\[
Z(T)=\sum_{M}e^{-\beta E_{M}}\]
they inherit its causal structure. Thus it is guaranteed that commutators
will vanish at spacelike separations on the boundary. In the large
$c$ limit, likewise the bulk commutator will vanish at spacelike
separations.

To extend the results of the previous section to the finite temperature
case the periodicity in the $\phi$ direction must be taken into account.
The supergravity correlators (and their boundary limits) may be obtained
via an image sum\[
\left\langle \phi(x)\phi(x')\right\rangle _{BTZ}=\sum_{n=-\infty}^{\infty}\left\langle \phi(x)\phi(x'+2\pi ne_{\phi})\right\rangle _{AdS}\]
where $e_{\phi}$ is a unit vector in the $\phi$ direction. Working
on the covering space, we see that whenever an image is light-like
separated from another point, there will be in general a non-vanishing
contribution to the commutator. In the range $|\delta t|<2\pi R$
the commutator will vanish at timelike separations on the boundary
as in the previous section. 

An exact CFT that exhibits the same behavior is simply a free massless
boson in two dimensions. The basic chiral correlators are functions
of $t/R+\phi$, so are not only periodic in $\phi$ and in imaginary
time, but also under $t\to t+2\pi R$. Therefore the corresponding
commutators will also be periodic in time. These will vanish for the
range of times $|\delta t|<2\pi R$, since the only dependence is
through $t/R+\phi$. This idea may be extended to non-chiral correlators
built out of products of chiral and anti-chiral factors, since each
factor behaves as above. 

In the following we will assume there exist CFT's dual to gravity
in $AdS_{3}$ that satisfy this criterion, that commutators vanish
at spacelike separations, and have the periodic structure in the timelike
direction noted above.%
\footnote{Note we do not expect the dual CFT to gravity in the BTZ background
obey the strict periodicity in the time direction found in chiral
conformal field theories. As we will see, timescales of interest for
us are typically much larger than $2\pi R$, and would not appear
in such theories.%
} With this in mind, the results of Section \ref{sub:Pure-AdS-Spacetime}
carry over to the finite temperature case.

\section{Finite Central Charge}

We have seen that at $c=\infty$ there is a simple relation between
the analytic and causal properties of the boundary correlators and
bulk causality, even in a nontrivial BTZ black hole background. Now
we wish to incorporate the effect of finite $c$ corrections. A prescription
for estimating the magnitude of such corrections was given in \citep{Hamilton:2007wj}.
The essential idea was to follow the above prescription to determine
approximate local bulk observables, but put a cutoff on the range
of the time integral, restricting to $\mathrm{Re}|\delta t|<t_{c}$.
In the present work, we realize this as a cutoff on the radius of
the disc in the $(t,\mathrm{Im}\phi$) plane, to preserve the isometries
of the analytically continued geometry. The timescale $t_{c}$ is
determined by examining when finite $c$ corrections can become of
comparable magnitude to the semiclassical boundary correlator. 

A useful discussion of these effects can be found in \citep{Barbon:2003aq,Barbon:2004ce,Barbon:2005jr}.
Initially the boundary correlator decays exponentially as $\exp(-\Gamma t)$
with $\Gamma\sim T_{H}$ , the Hawking temperature. Nonperturbative
effects, which we cannot hope to describe by some straightforward
modification of our quasi-local operators%
\footnote{Note that one might try to view CFT operators as corresponding to
bulk operators inserted in some kind of linear superposition of geometries,
represented as a sum over bulk topologies \citep{Maldacena:2001kr,Hawking:2005kf}.
Here we have in mind trying to reproduce correlators of such operators
via a local effective field theory on a background of fixed topology,
as advocated in \citep{Lowe:2006xm} , so we will not pursue these
other interpretations here.%
}, become of comparable magnitude when \begin{equation}
e^{-\Gamma t_{c}}\sim e^{-S_{bh}}\,.\label{eq:crittime}\end{equation}
With this cutoff prescription, the difference between using exact
finite $c$ CFT correlators and semiclassical large $c$ correlators
will be of order $e^{-S_{bh}}$. Moreover the correlators will have
the same analytic structure for light-like separations on the boundary,
which is the relevant property for determining when the commutator
of bulk operators is nontrivial.

The entropy of the BTZ black hole is \[
S_{bh}=\frac{2\pi r_{+}}{4G}=\frac{\pi r_{+}c}{3R}\,,\]
using the identification between Newton's constant and CFT central
charge $c=3R/2G$ \citep{Brown:1986nw,Strominger:1997eq}. For the
BTZ black hole, this yields the time scale\begin{equation}
t_{c}=\frac{2\pi^{2}cR}{3}\,.\label{eq:timesc}\end{equation}

We can now investigate regions of the spacetime, as shown in figure
\ref{fig:The-Penrose-diagram}, where the cutoff finite $c$ correlators
can be reproduced to good accuracy by some local gravity action. Let
us begin with point 1 in figure \ref{fig:The-Penrose-diagram}, a
point outside the horizon of the BTZ black hole. A bulk operator located
as such a point is represented by integration over a finite size disc
on the right CFT, according to \eqref{eq:bulkbound}. The cutoff is
irrelevant provided \begin{equation}
r>r_{c}=r_{+}+2r_{+}e^{-4\pi^{2}cr_{+}/3R}\,.\label{eq:rmin}\end{equation}
This provides us with an indication that AdS/CFT is reproducing a
stretched horizon as in the membrane paradigm picture of black hole
evaporation \citep{Thorne:1986iy,Susskind:1993if}. This radius corresponds
to a proper distance\[
ds=Re^{-2\pi cr_{+}/3R}\]
from the horizon (in the radial direction). For a large black hole,
this is shorter than a Planck length. In this case, we would expect
perturbative gravity interactions to prevent us probing such short
length scales with probes built out of gravitational fields. Hence
the effective thickness where one would expect to probe non-perturbative
effects should extend out to Planck scales. The cross-over, when the
free two-point computation yields a Planck scale stretched horizon,
happens when $r_{+}\sim G\,\log c$.

For points satisfying the bound \eqref{eq:rmin} the analyticity results
of the previous section will carry over, regardless of whether we
use the exact finite $c$ CFT correlator, or the semiclassical boundary
correlator. Therefore even at finite $c$ bulk correlators built to
the right of the light-sheet emanating from the points $r=r_{c}$
will display commutators exactly vanishing at spacelike separations.
The correlators will generically differ from the semiclassical correlators
by terms of order $e^{-S_{bh}}$, since for $t<t_{c}$ this is the
expected order of magnitude of the difference between semiclassical
boundary correlators and the exact finite $c$ correlators. This also
agrees with Page's results on the expected initial information outflow
from a black hole \citep{Page:1993wv,Page:1993df}. This picture is
what one expects from black hole complementarity \citep{Susskind:1993if}
- the region outside the stretched horizon behaves as if the black
hole was replaced by a hot membrane along the stretched horizon, with
physics in this region causal as usual. It is also worth noting effective
field theory formulated in the region $r>r_{c}$ (with appropriate
boundary conditions) will have a discrete spectrum, like that of the
CFT. If the region up to the horizon is included, the spectrum in
the bulk becomes continuous.

Now let us investigate what happens for bulk operators with large
time separations $t>t_{c}$, such as a correlator between points 1
and 3 in figure \ref{fig:The-Penrose-diagram}. Provided the points
are radii $r>r_{c}$, the cutoff will not be relevant, and the commutator
will vanish at spacelike separations as expected. However once one
point, e.g. point 1 enters the stretched horizon region ($r<r_{c}$)
this will no longer be the case. Now we expect a region of bulk spacetime
when the disc regions on the boundary intersect (supposing for simplicity
$\delta\phi=0$) near $t=t_{c}$. In this region the commutator will
be nontrivial, even when the bulk points are spacelike separated.
The magnitude of the commutator can be estimated by using the semiclassical
boundary correlator, and computing the bulk correlator with and without
the cutoff. This generically yields a magnitude of order $e^{-S_{bh}}$.
Again, this is consistent with Page's estimate of initial information
outflow in black hole evaporation \citep{Page:1993wv}. Since a large
BTZ black hole is an eternal black hole supported by an incoming flux
of radiation, this is also of the correct order of magnitude to describe
unitary evolution of the black hole \citep{Maldacena:2001kr}. We
conclude effective field theory will give accurate results for correlators
with small numbers of local operators, however if the number becomes
of order $e^{S_{bh}}$, corresponding to a measurement of the majority
of the radiation outside the black hole, then we expect to find special
correlators where the errors add coherently, and the expected error
will be of order $1$. 

One can play the same game with one point outside the black hole and
one point inside, for example points 1 and 2 of figure \ref{fig:The-Penrose-diagram}.
Now the cutoff prescription must be applied to expressions of the
form \eqref{eq:bulkrtlf}, with the left CFT mapped into the right
via the antipodal map \eqref{eq:antipode}. The analysis is essentially
the same as that of points 1 and 3 discussed above, when one point
sits inside the stretched horizon, except now the region of integration
on the boundary includes a pair of disc regions for the operator inside,
separated in imaginary time. The results above can be easily extended
to this case, and again we generically expect commutators of order
$e^{-S_{bh}}$. We do not expect all correlators with large numbers
of operators near point 3 to be correctly reproduced by effective
field theory due to expected errors of order $1$.

Finally we can consider the correlator between two points inside the
horizon, such as points 2 and 4 of figure \ref{fig:The-Penrose-diagram}.
When the points are spacelike separated, the boundary disc regions
can intersect, leading to a nontrivial commutator of magnitude $e^{-S_{bh}}$.
As discussed in \citep{Lowe:2006xm} effective field theory can still
be usefully formulated inside the horizon, since such small effects
are not operationally observable. This relies on the fact that measurements
inside the horizon have an intrinsic accuracy since they must be performed
before timelike geodesics hit the singularity (which happens in proper
time less than $\pi R/2$, shorter than the light-crossing time for
a large black hole). 

To sum up, the cutoff prescription can be used to determine in what
regions of spacetime, or for what class of correlators, effective
field theory can be expected to break down. The estimates of the magnitudes
of the finite $c$ effects lead to a bulk picture compatible with
unitary evolution of the black hole as described by the conformal
field theory. The essential new point is that effective field theory
cannot simply be maximally extended over a spacetime region as general
covariance would suggest. Rather effective field theory can only be
properly formulated on patches of spacetime where finite $c$ effects
are small, with some given intrinsic accuracy.

\section{Speculations on small black holes}

To achieve a satisfying resolution of the black hole information problem,
one would like to see that results similar to those described above
could be generalized to the case of small black holes in higher-dimensional
AdS, that actually evaporate away completely. One of the main difficulties
here is finding a simple way to identify these states in the CFT.
Nevertheless, let us assume the basic ideas carry over, and see what
picture emerges.

The bulk to boundary mapping for higher dimensional pure AdS spacetime
has been studied in \citep{Hamilton:2005ju,Hamilton:2006az,Hamilton:2006fh}.
Generalizing to eternal black hole states is expected to be conceptually
straightforward, but technically more complicated, with fewer explicit
expressions available. In higher dimensions, the cutoff prescription
will involve integrals over a sphere on the analytic continuation
of the boundary. Nevertheless, the basic analyticity arguments presented
above should generalize. 

For low temperatures, thermal AdS is the geometry giving rise to the
largest entropy. In this background, the boundary correlator will
oscillate with time, rather than having exponential falloff. Small
black holes will be a minority subset of this low temperature canonical
ensemble. Nevertheless, let us suppose we can construct suitable {}``chemical
potentials'' to filter out thermal AdS, and allow a set of small
black holes to dominate the modified ensemble. We will assume the
boundary correlator in this ensemble of small black holes exhibits
thermal behavior, namely\[
\left\langle \phi_{0}(t)\phi_{0}(0)\right\rangle \sim e^{-T_{H}t}\]
which should be reasonable if a quasistatic approximation can be applied
to the ensemble of CFT states allowing us to assume approximate ergodicity.
The timescale at which the semiclassical correlator receives corrections
of relative order $1$ can be estimated via the same argument as before.
This leads to

\begin{equation}
t_{c}\sim S_{bh}/T_{H}\sim1/T_{H}^{3}\mathrm{\, in\,4d}\label{eq:smalltc}\end{equation}
and is of order the information retention time \citep{Page:1993wv,Susskind:1993mu},
as discussed in \citep{Lowe:2006xm} for general dimensions.

When considering the expectation value of the commutator of local
operators, the same picture as described above will emerge. In particular,
the commutator of an operator behind the horizon and one outside can
become non-zero, and of order $e^{-S_{bh}}$ when the time separation
approaches the information retention time $t_{c}$. Likewise we expect
the commutator between an operator inside the horizon and a large
number of local operators outside at $t>t_{c}$ can have deviations
from the semiclassical result of order 1, if the operators are chosen
so that the errors add coherently. This corresponds to determining
the internal state of the black hole by measuring the Hawking radiation
in a manner compatible with unitarity.

The key difference with small versus large black holes is that they
are not supported by a flux of infalling thermal radiation, so will
eventually evaporate away to the dominant entropy state, that of thermal
AdS \citep{Hawking:1982dh}. The time-dependent geometry can be treated
by working in an adiabatic approximation. In low-dimensional examples,
the bulk-boundary map for time-dependent geometries has been studied
in \citep{Lowe:2008ra}. However time dependence means the $t_{c}$
that enters into the cutoff prescription will depend on what region
of the spacetime you are trying to describe with effective field theory.
Clearly far from the endpoint of evaporation, the cutoff time will
diverge, as expected for thermal AdS. On the other hand $t_{c}$ should
be chosen according to \eqref{eq:smalltc} to yield a description
of bulk physics in the vicinity of the black hole. In general $t_{c}$
will need to vary in some position dependent way to optimize the regime
of validity of a patch of effective field theory.

As emphasized above, the new feature of this construction is a prediction
of when effective field theory breaks down. This does not coincide
with the standard view that effective field theory should be valid
away from regions with large curvature invariants. Rather the region
of validity is determined by the class of observables and the region
of spacetime under consideration in accord with the cutoff prescription.

\section{Conclusions}

A method for constructing approximate local bulk operators from a
finite $c$ CFT has been studied. This enables exact CFT results to
be turned into predictions for experiments conducted in the bulk spacetime.
The analysis of the two-point function in pure AdS, and in a black
hole background indicate the results agree well with semiclassical
effective field theory when expected, and disagree in a manner compatible
with unitarity of quantum gravity. This provides an example of a direct
derivation from AdS/CFT of the information theoretic implications
of unitarity on effective field theory considered in \citep{Lowe:2006xm}.
This is an important step toward resolving the black hole information
paradox.

The cutoff procedure is necessarily ad hoc, and one might try to find
a better cutoff procedure which maximizes the region of validity of
effective field theory. We believe this is simply a way to parametrize
the inherent imprecision of quasi-local observables in a theory of
quantum gravity. A general discussion of such observables can be found
in \citep{Giddings:2005id}. A nice example that illustrates this
point is the quantum geometric generalization of de Sitter space (and
its associated CFT) considered in \citep{Guijosa:2003ze}. In this
example the continuous de Sitter space is replaced by a quantum geometry
that for many purposes behaves as a lattice of points. Clearly any
attempt to replace the exact observables on quantum geometry via effective
field theory on a continuous classical geometry will necessarily involve
some ad hoc approximation of the fundamental observables. We regard
the cutoff procedure to be a success if a large class of observables
can be reconstructed to a good approximation in some finite region
of spacetime, and the prescription described here appears to satisfy
that criterion.

An important open question is to go beyond the two-point functions
studied in the present work, and consider the effect of interactions.
These can be studied perturbatively in $1/c$. The expectation is
these corrections can be captured by local terms in the effective
action, and that the cutoff prescription will not be modified in a
substantive way, until back-reaction on the geometry becomes significant. 

The difference between the cutoff correlators and the semiclassical
correlators is typically very small, of relative order $e^{-S_{bh}}$.
As we have seen, these effects preserve causality for regions outside
the black hole, but nevertheless they appear to violate general covariance,
as they prevent us from maximally extending our effective field theory
over all regions of low curvature. It would be very interesting to
develop a set of observables (presumably non-local) sensitive to these
violations of general covariance in the vicinity of a black hole.
This can lead to new experimental probes of quantum gravity.

\begin{acknowledgments}
I thank Alex Hamilton, Daniel Kabat and Gilad Lifschytz for collaboration
on previous work and Leon Cooper for helpful discussions. This research
is supported in part by DOE grant DE-FG02-91ER40688-Task A.
\end{acknowledgments}
\appendix

\section{Coincidence limit of two-point correlator}

In this appendix we will examine the limit of \eqref{eq:integral}
when $X\to0$, $T\to0$ so that the smeared operators are centered
at the same point on the boundary, but still retain general radial
positions $Z_{1},Z_{2}$. The $i\epsilon$ prescription renders the
integrals well-defined, but it will prove convenient to take the limit
$\epsilon\to0$ first, and analytically continue in $Z_{1}$ and $Z_{2}$
to avoid the singularities that appear when $r_{1}=r_{2}$ at $\epsilon=0$.
First the integral over $z_{1}$ is performed, by evaluating the residue
at \[
z_{1}=\frac{r_{1}z_{2}}{r_{2}}\,,\]
(note the residue at $z_{1}=0$ vanishes for $\Delta\geq1$) and using
the definition of the Jacobi polynomial\[
P_{\Delta-1}^{(\beta,\gamma)}(x)=\frac{(-1)^{\Delta-1}}{2^{\Delta-1}(\Delta-1)!}(1-x)^{-\beta}(1+x)^{-\gamma}\frac{d^{\Delta-1}}{dx^{\Delta-1}}\left(\left(1-x)^{\Delta-1+\beta}(1+x)^{\Delta-1+\gamma}\right)\right)\,.\]
This gives the residue\[
Res=\frac{(\Delta-1)^{2}(-1)^{\Delta+1}}{2\pi^{3}R}\frac{r_{1}}{z_{2}}\left(\frac{r_{2}^{2}-r_{1}^{2}}{r_{2}}\right)^{1-2\Delta}\left(\frac{Z_{1}^{2}-r_{1}^{2}}{Z_{1}}\right)^{\Delta-2}\left(\frac{Z_{2}^{2}-r_{2}^{2}}{Z_{2}}\right)^{\Delta-2}P_{\Delta-1}^{(0,1-2\Delta)}\left(1-\frac{2r_{1}^{2}}{r_{2}^{2}}\right)\,.\]
This has a simple pole at $z_{2}=0$, so the integral over $z_{2}$
may be easily done. To perform the integral over $r_{1}$ we apply
a formula rediscovered by Askey in 1975 \citep{Askey1975}\begin{eqnarray}
_{2}F_{1}(a,b;c;x) & = & \frac{\Gamma(c)}{\Gamma(\mu)\Gamma(c-\mu)}\int_{0}^{1}dt\, t^{\mu-1}(1-t)^{c-\mu-1}(1-tx)^{\lambda-a-b}\,_{2}F_{1}(\lambda-a,\lambda-b;\mu;tx)\nonumber \\
 & \times & \,_{2}F_{1}(a+b-\lambda,\lambda-\mu;c-\mu;(1-t)x/(1-tx))\,,\label{eq:askey2}\end{eqnarray}
with $\mathrm{Re}c>\mathrm{Re}\mu>0$ and $|x|<1$. Note the left-hand
side is independent of $\mu$ and $\lambda$. We also need the relation
between the Jacobi polynomial and the Hypergeometric function\[
P_{n}^{(\beta,\gamma)}(x)=\frac{\Gamma(n+1+\beta)}{n!\Gamma(1+\beta)}\,_{2}F_{1}\left(n+\beta+\gamma+1,-n;1+\beta;\frac{1-x}{2}\right)\,.\]
Therefore after integrating over $z_{1},z_{2}$ and $r_{1}$ we arrive
at%
\footnote{The integral is valid for $|Z_{1}/r_{2}|<1$, however using the $i\epsilon$
prescription, we can continue $Z_{1}$ away from the real axis, and
use \eqref{eq:coinint} to define the integral throughout the complex
$Z_{1}$ plane.%
}\begin{equation}
\left\langle \phi(0,0,Z_{1})\phi(0,0,Z_{2})\right\rangle =\frac{\Delta-1}{\pi R}\int_{0}^{Z_{2}}dr_{2}\, r_{2}Z_{1}^{\Delta}(Z_{1}^{2}-r_{2}^{2})^{-\Delta}(Z_{2}^{2}-r_{2}^{2})^{\Delta-2}Z_{2}^{2-\Delta}\,.\label{eq:coinint}\end{equation}
Here we have simplified the resulting Hypergeometric function to an
algebraic function. Finally the integral over $r_{2}$ may be performed,
assuming $Z_{1}>Z_{2}$\[
\left\langle \phi(0,0,Z_{1})\phi(0,0,Z_{2})\right\rangle =\frac{Z_{1}^{2-\Delta}Z_{2}^{\Delta}}{2\pi R(Z_{1}^{2}-Z_{2}^{2})}\,.\]
This agrees exactly with \eqref{eq:ganal} in the same limit.

\section{Two-Point Correlator with one leg on the Boundary}

Here we will compute \eqref{eq:integral} in the limit $Z_{2}\to0$.
Without loss of generality, we can also set $T=0$. The integrals
over $r_{2}$ and $z_{2}$ are straightforward in this limit, giving\begin{eqnarray*}
\left\langle \phi(0,X,Z_{1})\phi(0,0,Z_{2})\right\rangle  & = & \frac{\Delta-1}{2\pi^{2}Ri}Z_{2}^{\Delta}\int_{C_{1}}dz_{1}\int_{0}^{Z_{1}}dr_{1}\frac{r_{1}}{z_{1}}\left(\frac{Z_{1}^{2}-r_{1}^{2}}{Z_{1}}\right)^{\Delta-2}\\
 & \times & \frac{1}{\left((X-i\epsilon+r_{1}z_{1})(X+i\epsilon-r_{1}/z_{1})\right)^{\Delta}}\,,\end{eqnarray*}
 and the limit $\epsilon\to0$ may be taken immediately. The analysis
proceeds in a similar way to the preceding appendix. As discussed
above, the contour $C_{1}$ encloses the point $z_{1}=r_{1}/X$ and
the origin. The residue vanishes at the origin, and at the point $z_{1}=r_{1}/X$
the residue may be written using a Jacobi polynomial. The integral
over $r_{1}$ may be performed using the formula \eqref{eq:askey2}
to give\[
\left\langle \phi(0,X,Z_{1})\phi(0,0,Z_{2})\right\rangle =\frac{Z_{1}^{\Delta}Z_{2}^{\Delta}}{2\pi R(X^{2}+Z_{1}^{2})^{\Delta}}\,.\]
This agrees exactly with \eqref{eq:ganal}.

\bibliographystyle{brownphys}
\bibliography{complement}

\end{document}